\newcommand{\jcapformat}{} 
\ifdefined\jcapformat
\documentclass[11pt,a4paper,pdflatex]{article}
\usepackage{jcappub}
\usepackage{float}
\else
\documentclass[usenatbib, twocolumn, nofootinbib]{aastex631}
\usepackage{color}
\fi

\usepackage{amsmath}
\usepackage{commands}
\usepackage{wrapfig}
\usepackage{comment}
\usepackage{gensymb}
\usepackage{multirow}
\usepackage{soul}
\usepackage{wrapfig}
\usepackage[normalem]{ulem}
\usepackage[mathlines]{lineno}
\usepackage{tikz}
\usetikzlibrary{shapes.geometric, arrows}
\usepackage{booktabs}
\usepackage{fontawesome}

\newcommand{\refresponse}[1]{\textcolor{black}{#1}}
\newcommand{\kp}[1]{\textcolor{blue}{[KP: #1]}}

\newcommand{\commenter}[1]{{}}

\newcommand{\srini}[1]{\textcolor{red}{[SR: #1]}}

\ifdefined\jcapformat
\newcommand{\apj}{ApJ}
\newcommand{\jcap}{JCAP}
\newcommand{\apjs}{ApJS}
\newcommand{\mnras}{MNRAS}
\newcommand{\aap}{A\&A}
\newcommand{\prd}{Physical Review D}

\fi

\newcommand{\diffusionmodel}{DDPM}
\newcommand{\mvirdef}{M_{500c}}
\newcommand{\massunits}{10^{14}}
\newcommand{\sigmaeight}{\sigma_{8}}
\newcommand{\agora}{Agora}

\newcommand{\Rvir}{R_{500c}}
\newcommand{\thetavir}{\theta_{500c}}
\newcommand{\clustermaskingrad}{3\thetavir}
\newcommand{\clustermaskingradformassiveclusters}{5\thetavir}
\newcommand{\minclustermaskingradinam}{10^{\prime}}
\newcommand{\sourcemaskingthreshold}{2}
\newcommand{\clustermaskingthreshold}{3 \times \massunits}
\newcommand{\patchsize}{6\degree \times 6\degree}
\newcommand{\pixres}{1.40625^{\prime}}

\begin{document}
\title{Learning Correlated Astrophysical Foregrounds with Denoising Diffusion Probabilistic Models}

\newcommand{\authora}{Karthik Prabhu}
\newcommand{\authorb}{Srinivasan Raghunathan}
\newcommand{\authorc}{Ethan B. Anderes}
\newcommand{\authord}{Lloyd E. Knox}

\ifdefined\jcapformat
\author[a,1]{\authora}
\author[b]{\authorb}
\author[c]{\authorc}
\author[a]{\authord}

\affiliation[a]{Department of Physics \& Astronomy, University of California, One Shields Avenue, Davis, CA 95616, USA}
\affiliation[b]{Center for AstroPhysical Surveys, National Center for Supercomputing Applications, Urbana, IL, 61801, USA}
\affiliation[c]{Department of Statistics, University of California, One Shields Avenue, Davis, CA 95616, USA}

\emailAdd{kprabhu@ucdavis.edu}

\else

\author[0009-0001-4135-8645]{Karthik Prabhu}
\affiliation{Department of Physics \& Astronomy, University of California, One Shields Avenue, Davis, CA 95616, USA}

\author[0000-0003-1405-378X]{Srinivasan Raghunathan}
\affiliation{Center for AstroPhysical Surveys, National Center for Supercomputing Applications, Urbana, IL, 61801, USA}

\author{Ethan B. Anderes}
\affiliation{Department of Statistics, University of California, One Shields Avenue, Davis, CA 95616, USA}

\author{Lloyd E. Knox}
\affiliation{Department of Physics \& Astronomy, University of California, One Shields Avenue, Davis, CA 95616, USA}

\correspondingauthor{K.~Prabhu} \email{karthikprabhu22@gmail.com}

\fi

\newcommand{\abstracttext}{
Extragalactic foregrounds -- most notably the Cosmic Infrared Background (CIB) and the thermal Sunyaev–Zel’dovich (tSZ) effect --- exhibit complex, non-Gaussian structure and correlations that can bias analyses of small-scale cosmic microwave background (CMB) temperature anisotropies. 
These foregrounds can introduce mode coupling ($\ell-\ell^{\prime}$ mixing) at small-scales (multipoles $\ell \geq 3000$) that mimic true lensing signals, thereby complicating analyses such as CMB lensing reconstruction. 
We present a novel approach to learn their full joint distribution using Denoising Diffusion Probabilistic Models (\diffusionmodel{}s)  trained on paired CIB–tSZ patches at 150 GHz, from the \agora{} suite of extragalactic sky simulations. While simulations like \agora, which are based on N-body calculations, can take thousands of CPU hours, \diffusionmodel{} can synthesize realistic CIB-tSZ patches that faithfully reproduce both auto- and cross-spectral statistics of the 2-point, 3-point, and 4-point correlation functions, in a matter of seconds. 
We further demonstrate matching pixel-value histograms and Minkowski functionals, confirming that conventional non-Gaussian benchmarks are also satisfied. 
This framework provides a powerful generative tool for forward-modeling correlated extragalactic foregrounds in current and future CMB analyses. 
Although we mainly demonstrate the joint modeling of tSZ and CIB at a single frequency, we also include examples of its extension to multiple frequencies, showing that the framework can learn the spectral energy distributions (SEDs) across different bands. 
While establishing \diffusionmodel{}s as a promising tool for addressing foreground contamination in next-generation CMB surveys, we also outline remaining challenges to their practical deployment in analysis pipelines, such as scaling to larger sky areas and reliance on the underlying cosmological and astrophysical assumptions in the simulations used for training. Our code and plotting scripts can be found on this \href{https://github.com/Karthikprabhu22/diffusion_model}{GitHub repo$^{\text{\faGithub}}$}.

}
\ifdefined\jcapformat
\abstract\abstracttext
\else
\begin{abstract}
\abstracttext
\end{abstract}
\fi


\maketitle

\section{Introduction} 
\label{sec_intro}

\commenter{
The millimeter wave observations of the cosmic microwave background (CMB), as one of the pillars of modern cosmology, have radically transformed the field of cosmological physics. ....
The secondary anisotropies of the CMB, arising due to interaction of the CMB photons with the matter in the Universe, are also becoming excellent probes of structure formation with the improvement in the quality of the small-scale measurements of the CMB. 
\srini{Not satisfied with these yet.}

In particular, gravitational lensing and the kinematic and thermal \sz{} (SZ) effects, have been providing crucial constraints on both $\lcdm$ and extensions to it including the constraints on $\sigmaeight$, the sum of neutrino masses $\summnu$ \citep{kaplinghat03, bocquet24b, ge24}, and also on astrophysics such as the detection of high redshift galaxy clusters \citep{...}, constraining the epoch of reionization \citep{..}, and others.
\srini{Really not satisfied at all.}
}

Millimeter-wave observations of the cosmic microwave background (CMB) have been central to the success of modern cosmology, providing a precise view of the early universe and placing stringent constraints on the parameters of the $\Lambda$CDM model. 
In addition to the primordial fluctuations, the CMB also carries imprints of its interactions with large-scale structure through secondary anisotropies, which are emerging as powerful probes of both cosmology and astrophysics. 
Foremost among these secondary effects are gravitational lensing and the kinematic and thermal \sz{} (SZ) effects, which have become critical tools for: probing structure formation; testing $\lcdm$ and constraining extensions to it via, for example, determinations of $\sigmaeight$ and the sum of neutrino masses $\summnu$ \citep{kaplinghat03, horowitz17, bocquet24b, madhavacheril23, ge24, qu24}; constraining the epoch of reionization \citep{reichardt21, gorce20, raghunathan24}; detecting high redshift galaxy clusters \citep{bleem15b, planck15-24, bleem20, hilton21, kornoelje25}; characterizing the thermodynamics of the hot gas and the impact of astrophysical feedback in galaxy groups \citep{schaan21, bigwood24, hadzhiyska24, reiduachalla25}.

However, the small-scale measurements of the CMB are significantly affected by contamination from instrumental noise, diffuse emission from dusty star-forming galaxies -- commonly referred to as the cosmic infrared background (CIB) -- and radio galaxies. 
Among these, CIB is particularly challenging to model since the diffuse emission is made up of multiple populations of galaxies over a wide range of redshifts, which contribute differently to each frequency band, introducing significant decorrelations between widely separated instrumental bands \citep{viero19}.
The CIB dominates temperature measurements at arc-minute scales and sets a floor for the current and future CMB temperature measurements \citep[see Figures~1 and 2 of][for example]{raghunathan23}. 

\commenter{
However, the small-scale measurements of the CMB also receive contribution from other undesired signals such as the instrumental noise, diffusion emission from dusty star-forming galaxies -- commonly referred to as the cosmic infrared background (CIB) -- and radio galaxies. 
Of these, CIB is the most difficult signal to be modeled \citep{...} which also sets a floor for the current and future CMB temperature measurements \citep[see Figs. 1 and 2 of][ for example]{crossilc}. 
This presents a significant hindrance in extracting information from the secondary anisotropies of the CMB. 
On the other hand, the presence of kSZ and tSZ signals, which can be highly non-Gaussian and correlated with large-scale structures, also introduce biases in CMB lensing \cite{crossilc, mh18, ferraohill17, ...}. 
Similarly, the presence of tSZ poses a strong challenge in robustly detecting the kSZ signal \cite{crossilc, ...}. 
}

While polarized CMB, which is relatively free from these contaminants, will be an important channel for future CMB lensing reconstruction, temperature-based CMB lensing will continue to add significant information over the next decade, and will dominate the small-scale lensing measurements in the future \citep{sehgal19}.
Similarly, robust measurements of small-scale temperature anisotropies will also be crucial for extracting information from kSZ and tSZ signals, and also for cross-correlations with galaxy surveys.

It is therefore essential to develop improved models of these contaminating signals.
In particular, it is important to jointly model the lensing, SZ, and CIB signals using correlated simulations of the extragalactic sky. 
Over the last few years, a few groups have produced correlated multi-tracer simulations \citep{sehgal19, stein20, omori24}. 
These simulations rely on high-resolution $N$-body simulations and ray-tracing, both of which are computationally expensive. 
As a result, producing on the order of thousands of such simulations is impractical and beyond the scope of any work in the near future.

Recently, several groups have successfully applied generative machine learning techniques for a wide range of astrophysical applications such as: generating correlated extragalactic foregrounds using Generative Adversarial Networks \citep{han21} and Normalizing Flows \citep{mebratu25}, reconstructing non-Gaussian CMB lensing maps \citep{floss24}, denoising weak lensing mass maps \citep{aoyama25}, and modeling dark matter and interstellar dust fields \citep{mudur22}.
In this work, we use the Denoising Diffusion Probabilistic Model framework (\diffusionmodel{}s) \citep{sohl-dickstein15, ho20, song20} to synthesize realistic samples of CIB and tSZ at $150\ \rm{GHz}$ consistent with the \agora{} simulations.
Using beyond-power spectrum statistics, specifically the one-point PDF, the collapsed bispectrum and trispectrum statistics, and the Minkowski functionals, we demonstrate that our generated maps accurately reproduce the non-Gaussian statistics of the \agora{} simulations. 
Furthermore, we show that our method can jointly model CIB and tSZ fields, enabling consistent generation of physically correlated map pairs.
While this work was in its final stages, \citet{aoyama25} published a similar work using the diffusion model framework to produce galaxy weak-lensing observables.

This paper is organized as follows. In \S\ref{sec_data}, we describe the \agora{} simulations of extragalactic foregrounds, focusing on CIB and tSZ maps at 150 GHz processed with point-source masking, and low-pass filtering. We detail the conversion of full-sky maps to $\patchsize$ patches for training, including data augmentation and normalization strategies. In \S\ref{sec_methodology}, we introduce the \diffusionmodel{} framework for generating simultaneous realizations of CIB and tSZ maps by learning their joint non-Gaussian distribution. In \S\ref{sec_results}, we demonstrate the performance of our model in reproducing key non-Gaussian features. We discuss the applications of \diffusionmodel{}s in extragalactic foreground modeling, alongside its limitations in \S\ref{sec_discussion} and finally conclude in \S\ref{sec_conclusion}.

\commenter{
As such, it is important to continue to investigate and model the contaminating signals \srini{Write this better}.
In particular, it is important to jointly model the lensing, SZ, and CIB signals using correlated simulations of the extragalactic sky. 
Over the last few years, a handful of groups have released correlated multi-tracer simulations \citep{sehgal19, stein20, omori24}. 
Generating these simulations necessitates N-body simulations as a starting point to produce halo catalogs and also ray-tracing simulations, which are both computationally expensive.  
As a result, producing order of thousands of such simulations is beyond the scope of any work in the near future. 

Recently, some groups have turned into generative machine-learning techniques to generate many number of correlated extragalactic realizations by learning from one of these existing simulations \citep{mmdl}.
In this work, we use the \diffusionmodel{} framework to generate fast realizations of \agora{} simulations. 
Using beyond-power spectrum statistics, specifically the collapsed bispectrum and trispectrum statistics, we show that our simulations can accurately reproduce the \agora{} simulations. 
We also show that our method can ... jointly model cib/tsz stuff ...
While this work was in its final stages, \citet{aoyama25} published a similar work using a diffusion model framework to produce galaxy weak-lensing observables. 

\srini{I think intro must be modified. Currently, the first three paragraphs are about CMB lensing. Better to simply remove these.}
\srini{Paragraph about CMB secondaries, and point to constraints from lensing, tSZ, and kSZ. Then mention the complexity of modelling the non-Gaussian foregrounds from CIB/tSZ for inference from future surveys. }
}

\commenter{
The gravitational lensing of the cosmic microwave background (CMB) offers a unique window into the large-scale structure of the universe, enabling precise constraints on several cosmological parameters such as the sum of neutrino masses ($\Sigma m_{\nu}$) \citep{kaplinghat03}, the dark energy equation of state($w$) \citep{dePutter09}, the amplitude of primordial non-Gaussianity ($f_{NL}$) \citep{hanson09}.

Gravitational lensing distorts primordial CMB fluctuations, imprinting non-Gaussian signatures through correlations between previously independent modes.
These lensing-induced non-Gaussianities encode a wealth of cosmological information, enabling reconstruction of the lensing potential power spectrum, $C_{\ell}^{\phi\phi}$, which traces the integrated matter distribution from $z\sim 1-4$ \citep{lewis06}. 
Lensing signals have been measured by several experiments, including Atacama Cosmology Telescope (ACT) \citep{qu24}, BICEP/Keck \citep{ade23}, {\it Planck} \citep{planck18-8} and South Pole Telescope (SPT-3G) \citep{pan23}.

However, this signal is entangled with non-Gaussian contamination from extragalactic foregrounds, such as the cosmic infrared background (CIB), Sunyaev-Zel'dovich (tSZ, kSZ) effect, and radio galaxies, which can cause biases in the lensing reconstruction. 
Common mitigation strategies include: 
1. Truncating temperature data at $\ell \geq 3000$ to avoid foreground dominated scales \citep{qu24}, 
2. Employing specialized estimators that leverages specific lensing signatures that are less prone to contamination by foregrounds \citep{osborne14}, or 
3. Multifrequency component separation under simplified spectral assumptions to reduce the impact of the foregrounds \citep{doohan25}. 
\srini{Several modified QE (gradient-cleaned QE, cross-ILC QE, etc.) have been introduced for these mitigations. Mick's paper actually does not do any mitigation.}
While these approaches reduce foreground biases, they come at the cost of discarding valuable information critical for lensing signal-to-noise (SNR) or rely on incomplete spectral models, limiting their applicability to next-generation surveys like CMB-S4 that would benefit from improved lensing measurements at high $\ell$s in temperature. 
\srini{Somewhere, we must mention that temperature adds info even for future surveys for CMB lensing. 
kSZ is all temperature. 
Otherwise, one can simply work with polarisation data.
}

Bayesian frameworks like MUSE \citep{millea21_muse} improve upon the traditional Quadratic Estimator (QE)--which uses only a first order
expansion of the lensed CMB fields in terms of the gradients of the lensing potential--by making use of all the available information in the fields.
As long as the simulation model is accurate, the MUSE analysis will remain unbiased \citep{ge24}.
However, the assumption of Gaussian model for foregrounds break down at small scales in temperature, where non-Gaussian foregrounds skew posterior sampling, causing biases in $C_{\ell}^{\phi\phi}$ as shown in \citep{doohan25}. 
Recent work with MUSE on polarization data \citep{ge24} avoids this issue by neglecting foregrounds, but temperature-dominated surveys critical for large-area lensing maps cannot ignore these contaminants. 

Recent advances in generative machine learning provide a promising path forward. 
Denoising Diffusion Probabilistic Models (\diffusionmodel{}s) \citep{sohl-dickstein15, ho20, song20} have emerged as a powerful tool for modeling complex data distributions. 
\diffusionmodel{}s excel in capturing non-Gaussian structures through a learned diffusion process that reverses a controlled noise‐injection schedule, iteratively denoising random noise into samples that can reproduce complex, higher‐order correlations present in the target distribution. 
This capability makes them uniquely suited towards Bayesian forward modeling approaches like MUSE, where generating accurate simulations is crucial. 
By integrating \diffusionmodel{}s into MUSE, we can replace the restrictive Gaussian priors with physically motivated distributions, enabling unbiased recovery on the lensing potential.

In this work, we present the first application of \diffusionmodel{}s to jointly model CIB and tSZ at 150 GHz, as a proof of concept. We train a \diffusionmodel{} on realistic paired CIB-tSZ patches from the \agora{} suite of simulations. We demonstrate that this approach can learn the joint-distribution of CIB and tSZ in the \agora{} simulations, and accurately reproduce the statistics of both the diffuse and the clustered components. Once trained, the model can synthesize new, independent patch pairs in seconds, enabling robust marginalization over foreground uncertainties. 
}


\section{Data}
\label{sec_data}

\subsection{Agora simulations}
\label{sec_agora}
The \agora{} suite of simulations \citep{omori24} provides a coherent framework for modeling correlated extragalactic foreground and secondary CMB anisotropies. 
The suite includes simulations of CMB lensing, thermal and kinetic \sz{} (tSZ/kSZ) effects, cosmic infrared background (CIB), and radio galaxies, all implemented within a unified lightcone. 
The simulations are produced by integrating dark matter halos and particles from the \textsc{Multidark-Planck2} (MDPL2, \citealt{klypin16}) $N$-body simulation. 
\agora{} \citep{omori24} self-consistently captures correlations between different observables, which is essential for studying biases in CMB lensing \citep{vanengelen13, madhavacheril18, sailer20, raghunathan23, maccrann23, doohan25}, kSZ \citep{raghunathan24}, and cross-correlations of these secondaries anisotropies with other tracers of large-scale structures \citep{fabbian19, omori23}.

In this work, we focus on two foreground signals, CIB and tSZ, that pose serious challenges for cosmological inferences such as CMB lensing, kSZ, and small-scale CMB temperature power spectrum analyses. 
The CIB and tSZ are non-Gaussian and also correlated with each other. 
This choice allows us to evaluate the performance of \diffusionmodel{}s on correlated, highly non-Gaussian extragalactic foregrounds before extending to more complex combinations. 
Throughout this work, we work with the simulated maps from \agora{} centered at the 150 GHz band using a bandpass similar to the South Pole Telescope SPT-3G experiment \citep{sobrin22}.
\agora{}'s tSZ signal map is generated by pasting halo profiles, calibrated using \textsc{Bahamas} hydrodynamical simulation \citep{mccarthy17, mead20}, at the halo locations.
The CIB signal map is produced by assigning stellar mass and star formation rate for each halo using the \textsc{UniverseMachine} \citep{behroozi19} code.

We use \agora{}'s CIB and tSZ simulations to generate both the training and validation samples. 
To this end, we extract a number of $\patchsize$ sky patches with $\pixres$ pixels from the full-sky HEALPix \cite{gorski05} map. 
We also apply a random rotation and flip to each patch to increase the number of training patches. 
While such augmentations introduce some correlation between derived samples, they still provide additional examples of localized structures in varying orientations, which improves model generalization without significantly biasing the learned statistical properties.
The sky patches are split into 80\% for \diffusionmodel{} training and 20\% for validation. 
We normalize the pixel intensities to [0,1] to stabilize training. 
However, unlike discriminative models, diffusion models do not rely on validation or test sets for early stopping or model selection, as their evaluation typically involves generating samples and comparing distribution-level statistics. 

Before extracting the sky patches, we also apply the following to the full-sky CIB and tSZ maps.
\begin{itemize}
    \item{ {\bf Point Source Masking}: We mask sources brighter than $\sourcemaskingthreshold\ {\rm mJy}$ using a single-pixel mask. 
    This threshold roughly corresponds to  $10\sigma$ detections in upcoming experiments \citep{cmbs4collab22}. 
    The masked pixels are set to zero, and the overall sky fraction removed is less than 1\%.
    While single-pixel masking is not realistic for actual data, note that the simulations used for \diffusionmodel{} training do not have an experimental beam and hence using a single pixel is sufficient for masking the source signal.
    }
    \item{{\bf Cluster Masking:} We mask locations of clusters with mass $\mvirdef \geq \clustermaskingthreshold$, approximately corresponding to $\sim 10\sigma$ detections for upcoming CMB surveys \citep{raghunathan22b}.
    Unlike point sources, the signal from galaxy clusters is not contained within a single pixel. 
    We therefore apply circular masks with radii ranging from $\clustermaskingrad$ to $\clustermaskingradformassiveclusters$ depending on the cluster mass, with a minimum masking radius set to $\minclustermaskingradinam$. 
    While this is an extremely conservative masking, the mask only removes $\sim3-4\%$ of the sky fraction, and thus we do not modify it further.
    Masked regions are inpainted with Gaussian random values with a mean and standard deviation corresponding to the entire map.
    In the above equation for the masking radius, $\thetavir$ corresponds to the angular extent of the cluster defined as $\thetavir = \Rvir / D_{A}$ where $\Rvir$, expressed in Mpc, is the physical radius of the cluster at which the mass density is 500 times the critical density of the Universe at the cluster redshift $z$ and $D_{A}$ is the angular diameter distance to redshift $z$ in Mpc.
    }
    \item{ 
        {\bf Low-pass filtering:} Finally, we low-pass filter the simulations to remove aliasing artifacts from scales $\ell \ge 7000$ while extracting the  patches.
    }
\end{itemize}


\begin{figure*}
    \centering
    \includegraphics[width=1.13\textwidth]{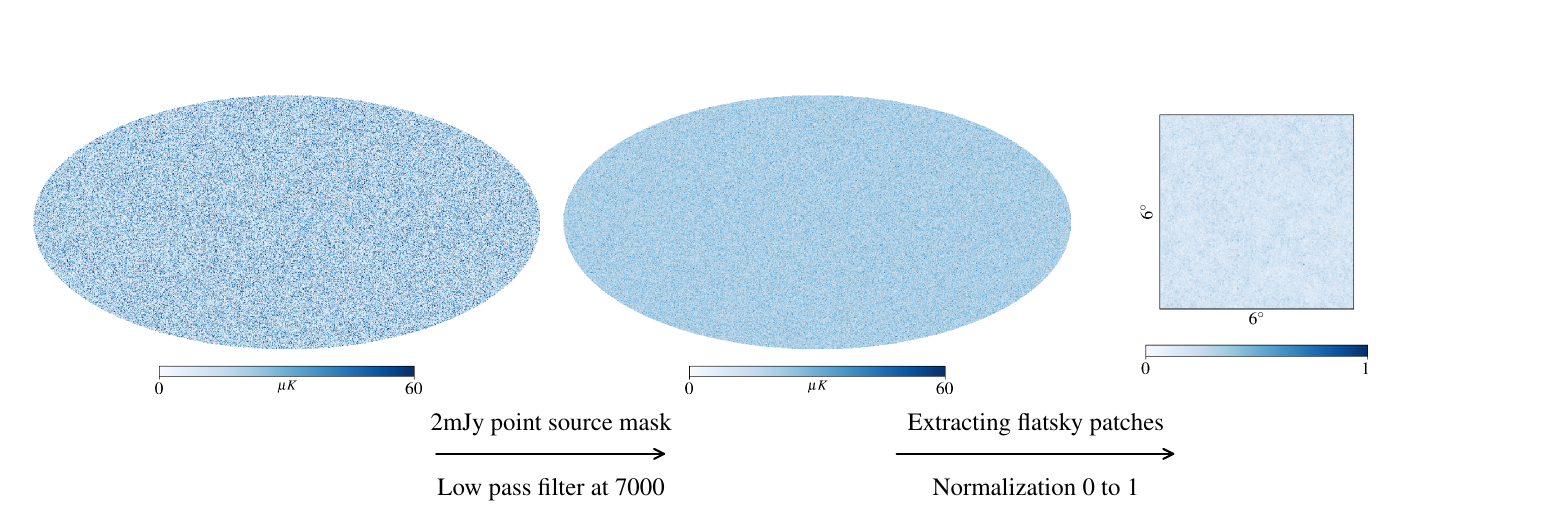}
    \caption{Schematic flowchart illustrating the processing pipeline of a full‐sky CIB map. The original full‐sky CIB map (left) is first processed by applying a $2\ \text{mJy}$ point source mask and a low pass filter at $7000$, yielding a filtered full‐sky map. Subsequently, $\patchsize$ patches are extracted from this processed map and min-max normalized to a range of 0 to 1.}
    \label{fig_map_processing}
\end{figure*}

\commenter{
\subsection{Cosmic Infrared Background}
\label{sec_cib}
\srini{I do not think this section is necessary.}
The Cosmic Infrared Background (CIB) is a diffuse extragalactic radiation field primarily originating from thermally re-radiated starlight absorbed by interstellar dust in the active star-forming galaxies at redshift z $\sim$ 1-2. 
This radiation dominates the far-infrared sky, emitting strongly at wavelengths between 100 $\mu\rm m$ and 600 $\mu\rm m$. 
This diffuse radiation field traces the integrated star formation history of the universe, and the large scale distribution of galaxies. 
Observations of the CIB have been made using infrared satellites such as IRIS \citep{depontieu14}, Herschel \citep{pilbratt10} and {\it Planck} \citep{planck18-4}, as well as ground-based CMB experiments detecting its subdominant contribution at lower frequencies \citep{viero19}. 
Therefore, building accurate CIB models is crucial for investigating contamination in these maps.

Spatially, the CIB exhibits significant clustering due to the large-scale structure (LSS) of the universe, as galaxies predominantly reside in dark matter halos. 
This clustering introduces non-Gaussian statistical features in the CIB maps, manifesting as correlated intensity fluctuations that cannot be fully described by power spectra alone. \kp{Expand more on this}
These non-Gaussian features correlate with the underlying matter distribution, introducing biases in the CMB lensing reconstruction where these sources mimic lensing-induced mode coupling \citep{george15}. 
This underscores the need for more advanced models, such as those leveraging dark matter halo catalogs \citep{}, to disentangle  the CIB's non-Gaussian imprint and mitigate biases in lensing analysis pipelines.

\subsection{Thermal SZ}
\label{sec_tsz}
\srini{I do not think this section is necessary.}
....
}


\section{Methodology}
\label{sec_methodology}
\subsection{DDPM Framework}
\label{sec_DDPM_framework}
Denoising Diffusion Probabilistic Models are a class of generative models that can learn complex data distributions and generate new samples from the learned distribution. Typically, a \diffusionmodel{} contains two processes: a \textbf{Forward process} which iteratively perturbs a highly structured data sample and gradually adds random noise over many steps, destroying the original structure; and a \textbf{Reverse process} that then learns to restore structure in the data by removing noise iteratively. Reversing such an entropy-increasing process normally requires an external influence, which in this case is provided by a neural network that encodes information about the structures and correlations present in the original data distribution. The reverse process can be seen as a time-reversed stochastic differential equation (SDE), where the neural network applies a time-dependent drift proportional to the score function ($\nabla \log p(\mathbf{x})$). Learning this score field is the essential core of \diffusionmodel{}s. 

\paragraph{Forward Process}
Let $\mathbf{x}_0$ be a data sample from the true distribution. The forward process defines a Markov chain that gradually perturbs $\mathbf{x}_0$ into standard Gaussian noise over $T$ steps with a fixed variance schedule.
$$q\left(\mathbf{x}_{1: T} \mid \mathbf{x}_0\right):=\prod_{t=1}^T q\left(\mathbf{x}_t \mid \mathbf{x}_{t-1}\right), \quad q\left(\mathbf{x}_t \mid \mathbf{x}_{t-1}\right):=\mathcal{N}\left(\mathbf{x}_t ; \sqrt{1-\beta_t} \mathbf{x}_{t-1}, \beta_t \mathbf{I}\right)$$
One can check that this formulation ensures that after long enough time, the final state $\mathbf{x}_T$ tends to a standard Gaussian distribution. We use $T=1000$ timesteps with a sigmoid schedule $\beta_t$ ranging from $\beta_1 = 10^{-4}$ to $\beta_{1000} = 0.02$.

\paragraph{Reverse process}
The reverse process is modeled by a neural network trained to approximate the distribution:
$$
p_\theta\left(\mathbf{x}_{t-1} \mid \mathbf{x}_t\right)=\mathcal{N}\left(\mu_\theta\left(\mathbf{x}_t, t\right), \beta_t\right),
$$
Our \diffusionmodel{} builds on the architecture of \citet{lucidrains22}\footnote{\url{https://github.com/lucidrains/denoising-diffusion-pytorch}}, which uses a U-net \citep{ronneberger15} architecture with self-attention mechanisms for the reverse process. 
U-Net is a convolutional neural network well-suited for image-to-image tasks. 
The U-net architecture consists of two major components: an \emph{Encoder} (contracting path) and a \emph{Decoder} (expanding path). 
The contracting path reduces the spatial resolution of the input images while capturing increasingly abstract features through a series of convolutional and downsampling operations. 
This allows the network to learn global context and structures. 
The expanding path then upsamples these features back to the original resolution, using transpose convolution layers. 
Crucially, the U-net also consists of \emph{skip connections} between the encoder and decoder blocks that allow fine-scale spatial information to be reintroduced during upsampling. 
This enables the network to localize and reconstruct fine-scale features, allowing for the recovery of small-scale features present in maps of CIB and tSZ. 
In our \diffusionmodel{}, we employ a series of four downsampling/upsampling blocks, and incorporate self-attention \citep{vaswani23} modules at intermediate resolutions to better capture long-range dependencies such as clustering. 

Directly optimizing the log-likelihood, $\sum_{i=1}^n\log p_\theta(\mathbf{x}^{(i)}_0)$, where $\mathbf{x}^{(1)}_0, \ldots, \mathbf{x}^{(n)}_0$ are the samples from $\mathbf{x}_0$, is intractable due to the complexity of marginalizing over all possible diffusion trajectories, which requires integrating over the full sequence of latent variables, $\mathbf{x}_{1:T}$.
Instead, we train the model by maximizing a lower bound on the negative log likelihood, known as the variational lower bound (VLB) or evidence lower bound (ELBO) as done in \citet[Eq. 3 of][]{ho20}.
The model is trained by minimizing an empirical estimate of this objective. Note that the objective does not directly incorporate any of the summary statistics described below. 

\commenter{
However, this objective can be simplified to a weighted mean squared error loss \cite{}. In our work, we adopt the velocity-prediction parameterization \citep{salimans22}, where the model learns to predict a latent variable. The Reverse Process trains the U-Net to predict the MSE loss on the velocity parameter $\epsilon_{\theta}$ as defined in \cite{salimans22}.
    \begin{equation}
        \mathcal{L}_\theta = \mathbb{E} \left[ || \mathbf{v} - \mathbf{v}_\theta\left(\mathbf{z}_t,t \right) ||^2\right], \quad \mathbf{v} \equiv \alpha_t \mathbf{z} - \sigma_t\mathbf{x}
    \end{equation}
This formulation improves numerical stability and aligns with the score-based interpretation: the network learns to estimate the direction in which the data becomes less noisy at each timestep.
}

\subsection{Summary Statistics}
\label{sec_summary_statistics}
We evaluate the fidelity of our trained model by using three sets of samples: hold-out patches from the \agora{} simulation, \diffusionmodel{}-generated samples, and patches from a Gaussian realization constructed to match the auto-power spectra of the CIB and tSZ as well as their cross-power spectrum. 
These Gaussian patches serve as a baseline for assessing the ability to capture non-Gaussian features beyond the two-point function.

To quantify the statistical precision expected in future experiments, we add S4-Ultra deep–like internal linear combination (ILC) residual noise based on \citet{raghunathan23} to both \agora{} and \diffusionmodel{} samples. 
This noise is included only when computing the variance (not the mean), allowing for an estimate of expected uncertainty without requiring explicit noise debiasing.

The first point of comparison is the angular power spectrum, which quantifies scale-dependent correlations in the maps. 
We compute it using a binned flat-sky power spectrum estimator over a range of $300<\ell<4000$, in bins of width $\Delta\ell = 60$.
For the \diffusionmodel{}-generated samples, we initially observe a slight deficit in the power spectrum amplitude, dominant in the Poisson part of the power spectrum. 
We attribute this to the difficulty of capturing rare, high-intensity pixels—an established limitation of diffusion models \citep{stamatelopoulos2024}. 
To mitigate this, we apply a post-training rescaling of the pixel intensities.
\refresponse{Specifically, we multiply each \diffusionmodel{} sample by a single global factor: the ratio of the standard deviation of all the \agora{} samples to that of all the generated samples. 
For the CIB maps, we find a rescaling factor of $1.0328$, while for the tSZ maps, we obtain $1.1425$.}
This post-training adjustment proves remarkably effective: it restores agreement with the target power spectrum and also substantially improves the performance of the model across all subsequent summary statistics, without having to retrain the model.

The more interesting metrics are those that quantify the correlations beyond the power spectrum. 
We looked at several benchmarks of non-Gaussianity, including physically motivated statistics like the bispectrum and the trispectrum, which are sensitive to lensing-induced mode coupling but are expensive to compute. We also examine more traditional measures such as the Minkowski functionals and normalized pixel intensity histograms. 
We discuss each of them below.

\begin{itemize}
    \item[(1)]{{\bf Pixel Intensity Histograms}:
Normalized histograms of pixel values probe the one-point probability density function (PDF). 
We compute normalized histograms for 200 samples of CIB and tSZ from test samples and \diffusionmodel{} generated samples. The histograms are constructed using 1000 bins over a range [0 $\mu K$, 100 $\mu K$] for CIB and [–100 $\mu K$, 0 $\mu K$] for tSZ. The histograms are smoothed using a Gaussian kernel with $\sigma = 1$ to suppress noise. } 

\item[(2)]{{\bf Minkowski Functionals:}
Minkowski Functionals \citep{schmalzing95}  characterize the morphology of excursion sets -- defined as the pixels with intensities above a certain threshold $\nu$.
Typically, there are three functionals defined: 
$M_0(\nu)$: Area fraction (cumulative distribution function), 
$M_1(\nu)$: Total boundary length of excursion sets, and 
$M_2(\nu)$: Euler characteristic (number of connected components minus holes),
across a range of intensity thresholds $\nu$.
We use the \citet{boelens21} software package to compute these quantities for threshold intensities $\nu \in [0, 1]$ on normalized maps. }

\item[(3)]{{\bf Bispectrum and Trispectrum:} 
Although one could compare all higher-order moments, the three-point (bispectrum) and four-point (trispectrum) correlation functions are the most important for lensing analyses. 
For computational efficiency, we restrict our analysis to the equilateral configurations of the bispectrum and trispectrum which can be estimated from the skewness and kurtosis of the harmonic band-filtered maps, as done in \citet{lee24}. We compute these moments for the sum of CIB, tSZ, and S4-Ultra deep–like noise. We also evaluate these moments individually for CIB and tSZ as well as their cross moments, which can be found in Appendix~\ref{appendix_moments}. We demonstrate the ability of \diffusionmodel{}s to accurately capture these moments by comparing to Gaussian samples where we expect these moments to be consistent with zero.}
\end{itemize} 

\section{Results}
\label{sec_results}
We present the \diffusionmodel-based simulations in this section and compare them to the original \agora{} simulations. 
We start with basic visual inspections in \S~\ref{sec_visual_inspection} and then compare the tSZ profiles of haloes between the two simulations using a stacking approach.
Next, we perform thorough quantitative comparisons using the summary statistics described above in \S~\ref{sec_summary_statistics}. 
These include the power spectra comparisons, and also the ones that capture the non-Gaussianity in the data such as the pixel intensity histograms, Minkowski functionals, and higher-order correlation functions, namely the bispectrum and the trispectrum.

\subsection{Visual Inspection}
\label{sec_visual_inspection}
\begin{figure}[!htbp]
    \centering    \includegraphics[width=\linewidth, keepaspectratio]{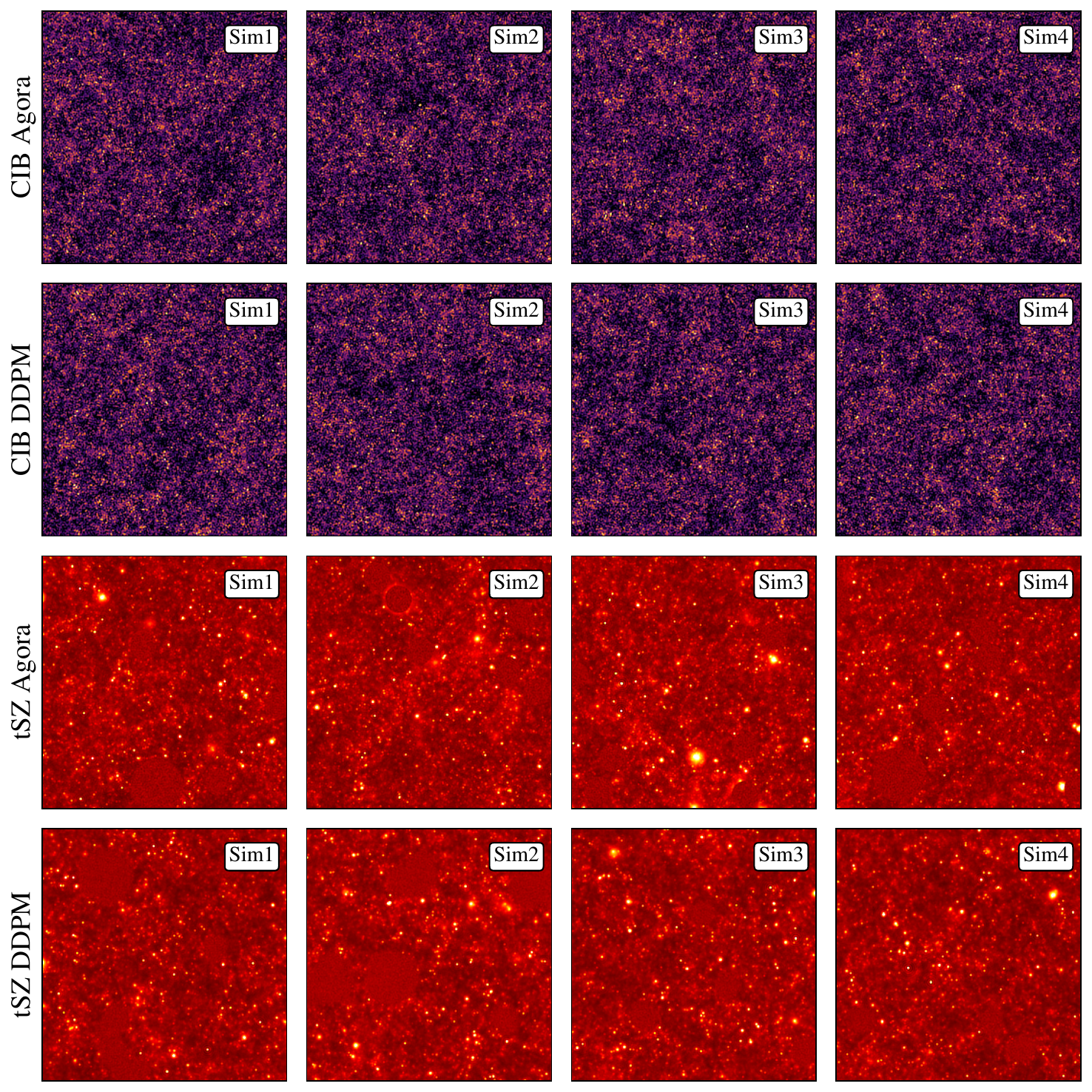}
    \caption{
    Randomly selected patches of the CIB (top two rows) and tSZ (bottom two rows) at 150 GHz. Each column shows a different simulation index. The first and third rows show samples from the \agora{} simulation, while the second and fourth rows show corresponding outputs from the \diffusionmodel{}. The \diffusionmodel{} closely matches the visual appearance of the training data, reproducing cluster masks, point sources, and even characteristic ringing artifacts from low-pass filtering.
    }
    \label{fig_examples}
\end{figure}

As a preliminary check, we examine a few randomly selected CIB-tSZ patch pairs from the \agora{} simulation and compare them to samples generated by the trained \diffusionmodel{} in Figure~\ref{fig_examples}. 
Visually, the \diffusionmodel{} samples are nearly indistinguishable from the \agora{} patches; without labels it would be difficult to tell them apart. The \diffusionmodel{} faithfully reproduces large circular cluster masks with similar morphology and statistical properties. It also captures isolated bright pixels corresponding to point sources in both CIB and tSZ maps, appearing at frequencies consistent with the training data. Due to the low-pass filtering applied during preprocessing, these sources sometimes produce characteristic ringing artifacts in their vicinity. The \diffusionmodel{}-generated samples also show similar bright spots with surrounding ripples akin to those present in the \agora{} patches.

\subsection{Stacked tSZ profiles}
\label{sec_tsz_stacks}

\newcommand{\tszcutoutsize}{31^{\prime} \times 31^{\prime}}
Besides the visual inspections mentioned above, we compare the profiles of the tSZ signals from \diffusionmodel{} with \agora. 
To this end, we use the tSZ simulations and extract $\tszcutoutsize$ cutouts centered at the locations of pixels $i$ that fall within a given SNR threshold (A,B) defined as $A\sigma_{T} < | T_{i} - \bar{T} | \le B\sigma_{T}$, where $\bar{T}$ and $\sigma_{T}$ represent the mean and the standard deviation of the simulated tSZ map.
Specifically, we pick regions in the following SNR ranges: (a) $5-10\sigma$, (b) $10-20\sigma$, and (c) $\ge 20\sigma$.
Note that the simulated maps are free from instrumental noise, and hence the noise corresponds to the tSZ-confusion noise \citep{raghunathan22b} from the other haloes present in the map.
The number of haloes picked from \agora{} and \diffusionmodel{} simulations agree within 5-10\% for the first two cases, but the difference is slightly higher ($\sim 19\%$) for the high SNR case. 
The larger discrepancy for the high SNR case arises from the difficulty in properly capturing the statistics of the massive clusters with the patch size used for training. 
We discuss this further in the subsequent sections.
\refresponse{
Once we collect the cutouts of haloes, we compute the stack in each SNR bin as
\begin{equation}
\hat{S}(\theta_x, \theta_y) = \frac{ \sum_{i}  w_{i} S_{i}(\theta_x, \theta_y)}{ \sum_{i} w_{i}}
\end{equation} 
where the index $i$ runs over all the clusters in a given SNR bin defined above and we apply uniform weights $w$ for all the clusters. Here, $\theta_x$ and $\theta_y$ are Cartesian angular coordinates with the origin at the cluster center. 
In Figure~3, we present the radial profile of the stacked tSZ signals, $\hat{S}(\theta_x, \theta_y)$, as a function of radius $\theta = \sqrt{\theta_{x}^2 + \theta_{y}^{2}}$ from the cluster center. 
To compute the radial profile, we average the stacked map $\hat{S}(\theta_x, \theta_y)$ over all pixels with coordinates $(\theta_x, \theta_y)$ such that $\sqrt{\theta_x^2 + \theta_y^2} \in (\theta_b, \theta_b + \Delta \theta)$, where $(\theta_b, \Delta \theta)$ define the edges of the radial bin.
We use 10 linearly spaced bins with $\Delta \theta = 1^{\prime}$ and $\theta_{\rm max} = 10^{\prime}$.}
In the figure, black corresponds to the radial profile from \agora{} and blue corresponds to \diffusionmodel{} simulations. 
In the top panels, we also show the difference between \agora{} and \diffusionmodel{} in yellow, and in the bottom panels, we present the ratio. 
As evident from the figure, radial profile of the tSZ signals from \diffusionmodel{} agrees with the ones from \agora{} to within 8\%. 
We also note that the radial profiles from both \agora{} and \diffusionmodel{} simulations do not reach zero even at $\theta \ge 10^{\prime}$ from the cluster center. 
This is due to the 2-halo term of the tSZ signal \citep{hill18} dominating over the 1-halo term at larger distances from the cluster center, and it is being reproduced in \diffusionmodel{} simulations to within 5\% when compared to \agora.

\begin{figure*}[!htbp]
    \centering
    \includegraphics[width=1\textwidth, keepaspectratio]{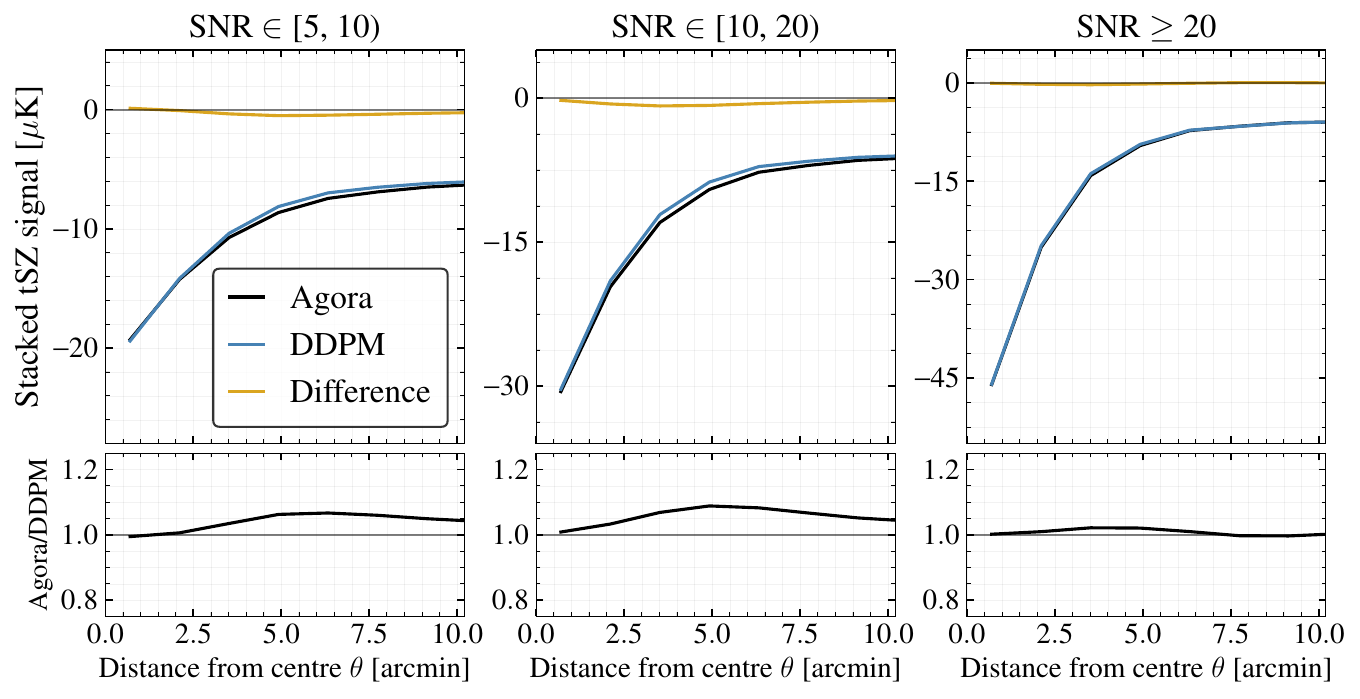}
    \caption{Radial profiles of stacked tSZ signals in three SNR bins. The black curves correspond to \agora{} and the blue to \diffusionmodel{} simulations. We stack roughly 260,000 in the first SNR bin $\in [5, 10)$; 60,000 in the second bin $\in [10, 20)$; and 3,900 in the third bin ${\rm SNR} \ge 20$. In the top panels, we show the difference between \agora{} and \diffusionmodel{} profiles in yellow, and present the ratio of the two as a function of distance from the center in the bottom panels. We find that the \diffusionmodel{} simulations agree with \agora{} simulations to within $8\%$. In the top panels, the profiles do not reach zero even at larger radii because of the 2-halo term.}
    \label{fig_tsz_stacks}
\end{figure*}

\subsection{Power Spectra comparison}
\label{sec_power_spectra_comparison}
\begin{figure*}[!htbp]
    \centering
    \includegraphics[width=\textwidth, keepaspectratio]{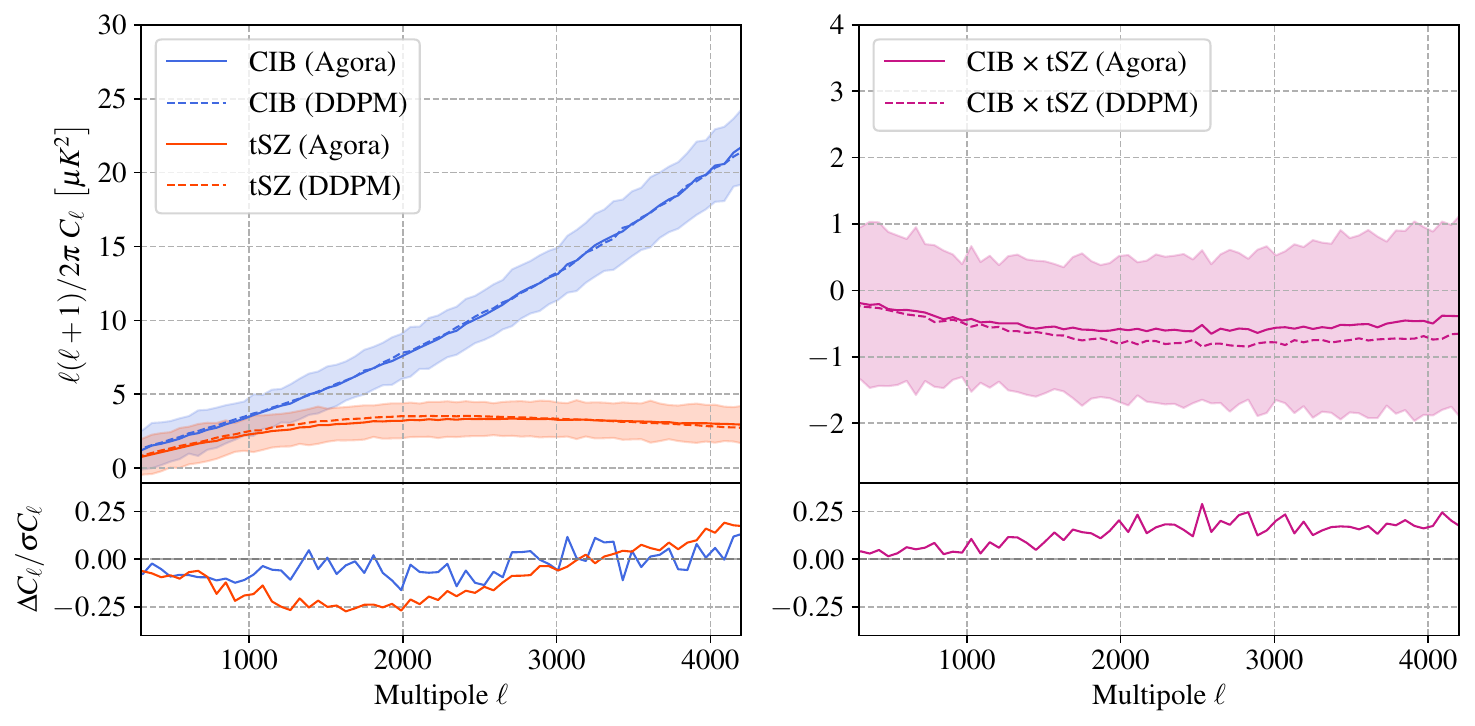}
    \caption{
    Comparison of angular power spectra between \diffusionmodel{} samples and test maps for both auto- and cross-correlations. In the top panel, the auto-power spectra $C_\ell$ for the CIB and tSZ components (blue and orange, respectively) and the cross-power spectrum $C_\ell^{\mathrm{CIB}\times\mathrm{tSZ}}$ (pink) are shown for both \diffusionmodel{} (solid lines) and test samples (dashed lines), across multipoles $\ell = 300$–$4200$. Shaded bands represent the $1\sigma$ sample variance from the test set. In the bottom panel, residuals between the test and \diffusionmodel{} power spectra, expressed as $(C_\ell^{\mathrm{Agora}} - C_\ell^{\mathrm{DDPM}})/\sigma(C_\ell^{\mathrm{Agora}})$ are shown.
}
        \label{fig_power_spectra_comparison}
\end{figure*}

Next, we evaluate the auto- and cross- angular power spectra of 200 randomly selected CIB and tSZ patch pairs generated by the \diffusionmodel{} and compare them to patches from the \agora{} simulation. 
\refresponse{The results of this comparison are shown in Figure ~\ref{fig_power_spectra_comparison}}. 
The CIB exhibits a Poisson-dominated spectrum, where $D_\ell$ increases as $\ell^2$, consistent with a model with Poissonian shot noise term and a continuous clustering term \cite{bond91}. 
In contrast, tSZ shows a peak structure due to the characteristic angular size of galaxy clusters, set by their halo pressure profiles. 
After applying variance rescaling as described in Section~\ref{sec_summary_statistics} to the \diffusionmodel{} outputs, to match the true variance, we find that the generated power spectra match the ground truth within 25\% of the sample variance across all multipoles. 
We find that the CIB power spectrum matches well even without rescaling, while the tSZ spectrum requires rescaling to correct for the underestimation of its extreme-valued pixels. 
The cross-spectrum between the CIB and tSZ, which probes the spatial correlation between dusty star-forming galaxies and the hot gas traced by tSZ, is also successfully reproduced by the \diffusionmodel{}, well within the expected sample variance.

\subsection{Pixel Intensity Histograms}
\begin{figure*}
    \centering
    \includegraphics[width=\textwidth, keepaspectratio]{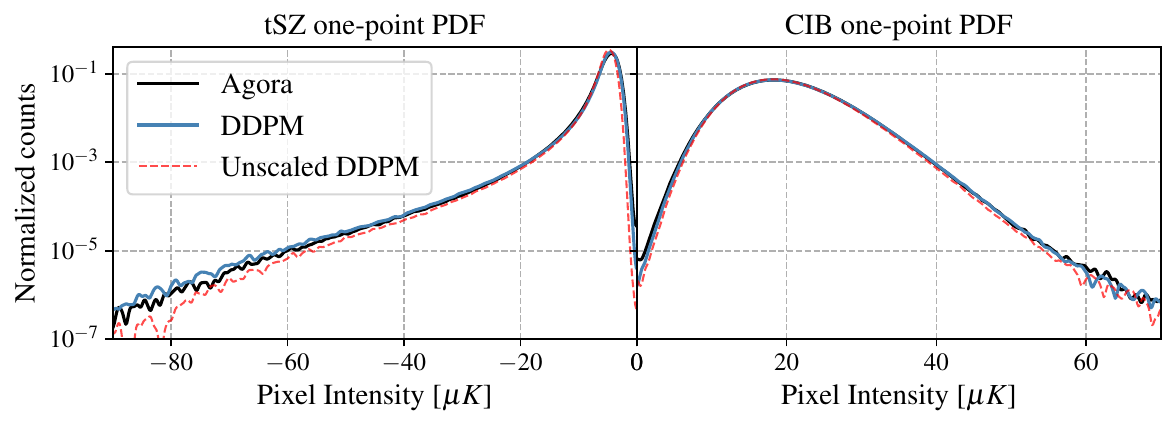}
    \caption{Comparison of one-point PDF: Pixel intensity density estimates for test samples (black) and \diffusionmodel{} samples (blue) using histograms. The pixel values were binned into linearly spaced bins, and the normalized histograms are smoothed with a 1D Gaussian filter ($\sigma=1$) to suppress the high frequency fluctuations near the tails. }
    \label{fig_histogram_comparison}
\end{figure*}

Figure~\ref{fig_histogram_comparison} presents the pixel intensity distributions for the CIB and tSZ maps. The \diffusionmodel{}-generated maps (dashed lines) closely follow the distributions of the \agora{} simulations (solid lines), especially near the mode. This confirms that the diffusion model successfully captures the bulk of the one-point statistics across both components.

\refresponse{
However, there are discrepancies present in the distribution tails. For the CIB, we observe a slight underproduction of high-intensity pixels, with a mild shortfall beyond intensities of $\sim 60\ \mu K$. 
For the tSZ, the \diffusionmodel{} underestimates the occurrence of strong decrements (large negative values), which correspond to massive clusters, as seen by the suppressed tails in the unscaled \diffusionmodel{} pixel histogram. These deviations highlight the challenge in recovering rare, high-amplitude fluctuations that drive the heavy tails of these distributions.
}

\subsection{Minkowski Functionals}
\begin{figure*}
    \centering
    \includegraphics[width=\linewidth, keepaspectratio]{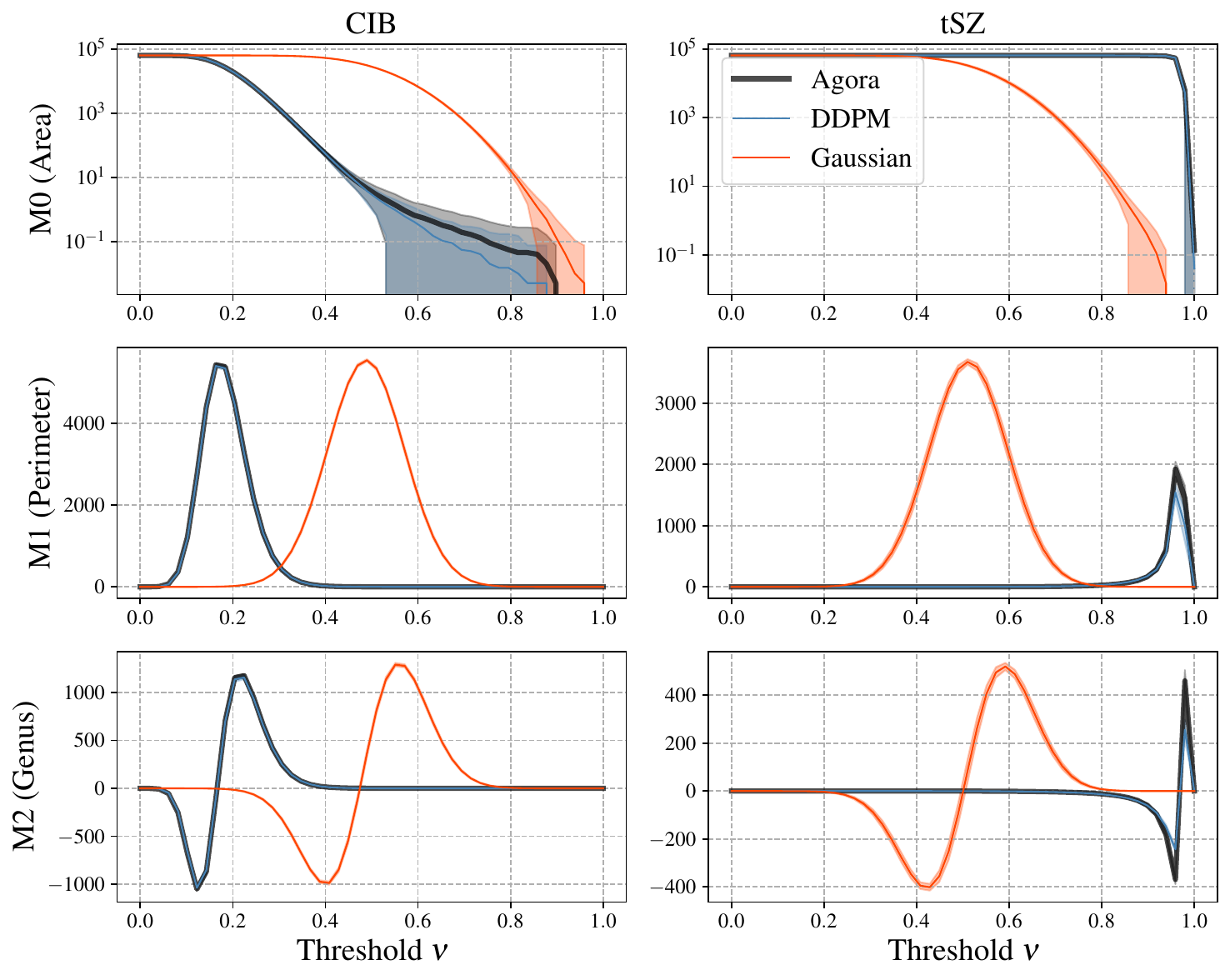}
        \caption{Minkowski functionals $M_0$ (Area), $M_1$ (Perimeter), and $M_2$ (Genus) as a function of intensity threshold $\nu$ for cosmic infrared background (CIB, left) and thermal Sunyaev-Zel’dovich effect (tSZ, right). We compare real maps from \agora{} (black), DDPM-generated samples (blue), and Gaussian simulations (orange). Shaded regions denote standard deviations across realizations.
        }
    \label{fig_minkowski}
\end{figure*}

Figure~\ref{fig_minkowski} shows the Minkowski functionals $M_0$, $M_1$, and $M_2$ as a function of threshold $\nu$ for both the CIB (left) and tSZ (right) maps. The DDPM samples (blue) closely match the real \agora{} data (black) across all three statistics, demonstrating the model’s ability to reproduce both global and topological features of the fields. For the tSZ channel, there is a small mismatch between the DDPM and \agora{} curves in $M_1$ and $M_2$, particularly near high-threshold regions where signal is sparse. However, these deviations are minor compared to the significant differences observed with the Gaussian simulations (orange). The Gaussian maps fail to capture the heavy tails and non-Gaussian features present in the real and DDPM-generated maps, resulting in smooth, symmetric $M_1, M_2$ curves centered at 0.5. 

\subsection{Bispectrum and Trispectrum}
\label{sec_bispectrum_trispectrum}
\begin{figure*}
    \centering
    \includegraphics[width=\linewidth, keepaspectratio]{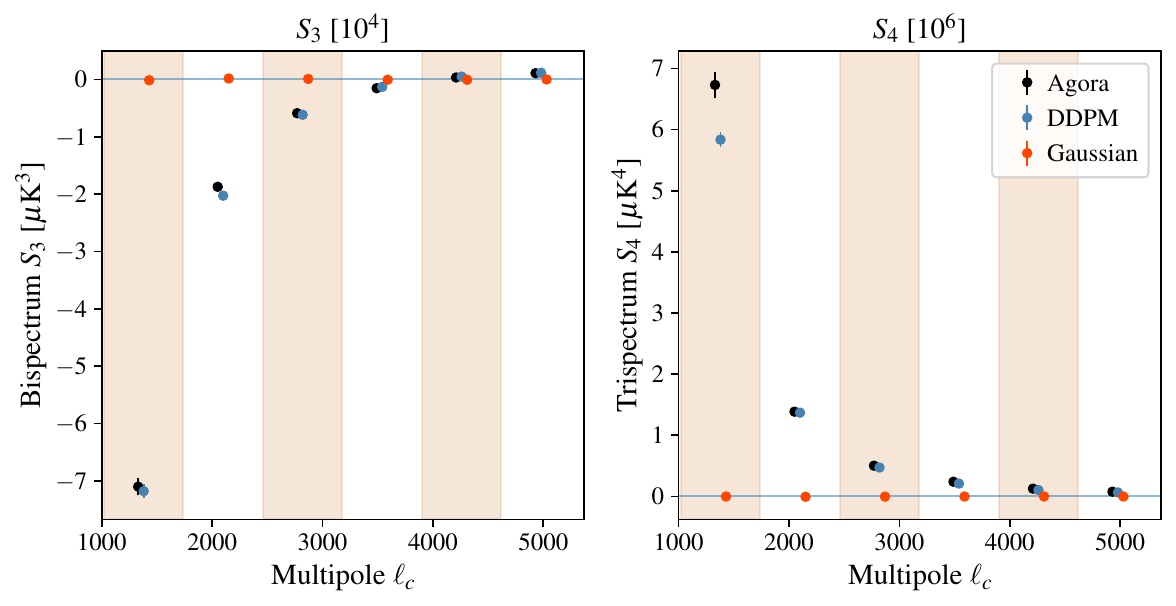}
        \caption{Comparison of bispectra and trispectra of the sum of CIB and tSZ signals and noise as a function of band center, $\ell_c$ . Each panel compares the mean and standard error of the statistics across 800 patches from three models, Gaussian simulations, \agora{} simulations, and \diffusionmodel{} samples, for S4-Ultra deep–like instrumental noise levels.
        }
    \label{fig_moments_joint}
\end{figure*}

In Figure~\ref{fig_moments_joint}, we present the estimated collapsed equilateral bispectrum ($S_3$) and trispectrum ($S_4$) of the sum of CIB, tSZ, and S4-Ultra deep–like instrumental noise, evaluated across a range of harmonic band centers $\ell_c$. We compare the results from \agora{} simulations, \diffusionmodel{} samples, and Gaussian realizations. As expected, the Gaussian samples remain consistent with zero across all $\ell_c$, serving as a baseline. The \diffusionmodel{} successfully recovers the non-zero skewness and kurtosis associated with the foregrounds in the \agora{} simulations. However, minor differences are present between the trispectrum of \agora{} and the \diffusionmodel{} generated samples at low multipoles. 
A more detailed breakdown of third- and fourth-order auto- and cross-moments between the CIB and tSZ components, under SPT-3G, S4-Wide, and S4-Ultra deep noise levels, is provided in Appendix~\ref{appendix_moments}. 

\refresponse{
We investigated if the underproduction of extreme-value pixels could be the source of the large-scale discrepancy between \agora{} and \diffusionmodel. 
To this end, similar to the stacking exercise in \S~\ref{sec_tsz_stacks}, we select pixel locations satisfying with ${\rm SNR} \ge 10\sigma$, corresponding to the extreme-value pixels, and use a 2D Gaussian fitting approach to determine the size of the haloes. 
Next, we mask the locations of these high SNR signals in both \agora{} and \diffusionmodel, and recalculate the bispectrum and trispectrum statistics. 
While masking these extreme signals reduces the non-Gaussianities suppressing both the bispectrum and the trispectrum, it does not have a noticeable impact on the differences between \agora{} and \diffusionmodel. 
It is important to note here that we do not retrain the \diffusionmodel{} after masking the extreme-value pixels and leave a more careful investigation of this discrepancy to a future work. 
}

\section{Discussion}
\label{sec_discussion}

In this section, we discuss some of the applications we envision for \diffusionmodel{}s in extragalactic foreground modeling, outline some practical challenges to their deployment in analysis pipelines, and compare \diffusionmodel{}s with other generative modeling approaches.

\commenter{\sout{A second challenge concerns the treatment of masked regions.
Our training pipeline involves masking point sources and clusters present in the individual extragalactic foregrounds, which is possible because we have access to these components via the simulations. 
For example, we use single-pixel masks for point sources and fixed-radius circular masks for clusters, with inpainting via Gaussian-distributed random fields. 
In observational data, however, the individual signal, noise and foreground components are not available independently.
While some sources are well-resolved and can be reliably masked, some can be ambiguous. Training diffusion models to handle masked and unmasked regions jointly, potentially through conditional diffusion models, remains an open direction for future work.}}

\subsection{Applications}
We imagine two primary applications for \diffusionmodel{}s in CMB analysis: (1) as a rapid generator of simulations that can be used in downstream analysis pipelines, and (2) as a probabilistic model in a Bayesian inference engine. We discuss them below.

\begin{itemize}
\item[(1)]{Correlated multi-tracer simulations such as \agora{} or Websky rely on computationally expensive $N$-body simulations and ray-tracing, and often require several thousands of CPU hours to generate a single realization. 
\diffusionmodel{}s on the other hand, once trained on a few thousand patches from these simulations, can generate realistic samples with accurate summary statistics in a matter of seconds, as demonstrated in this work. These models can rapidly generate large ensembles of simulations to test for possible biases and to build accurate covariance matrices.}

\item[(2)]{The trained \diffusionmodel{} can be incorporated into Bayesian inference engines like MUSE \cite{millea21_muse}. 
Within such a framework, the \diffusionmodel{} can be used in two ways: either as part of the data model, where it can replace simplistic Gaussian priors with physically motivated, complex foreground distributions, or in the forward simulator, where it generates synthetic skies that propagate through the full pipeline to correct for biases in the inference process.}
\end{itemize}

\subsection{Real-world challenges}
\label{sec_real_world_challenges}

\noindent
{\it Increasing the size of the simulations:} 
Perhaps the most significant challenge in applying \diffusionmodel{}s to real-world analysis tasks is their extension to larger sky areas.
The number of available training samples decreases as the patch size increases, since only a handful of correlated multi-tracer simulations currently exist. 
Moreover, the computational cost of training these models scales typically as $\mathcal{O}(N_{\rm pix}^2)$ or higher.
In this work, we train on patches of size $\patchsize$, which lets us extract about 4000 training examples (with data augmentation) from the \agora{} simulation.
This was the largest patch size for which we were able to train a model reliably. We experimented with larger patches of size $10^\circ \times 10^\circ$, but observed a significant degradation in sample quality, likely due to insufficient training data.
However, most of the current surveys span sky areas at least 10 times larger. 
Extending \diffusionmodel{}s to generate larger sky areas thus remains a challenge. 
One promising direction is the Patch Diffusion framework proposed in \cite{wang23}, which we leave for future work.\\

\noindent
{\it Reproducibility of extreme-value pixels:} 
A consequence of the above limitation in the number of available sky patches for training, is the difficulty in capturing the statistics of extreme-value pixels. 
In this work, we masked sources brighter than $\sourcemaskingthreshold\ {\rm mJy}$ and clusters with mass $\mvirdef \ge \clustermaskingthreshold$. 
Since both the source number counts and the halo mass function are a steep function of the source flux or the cluster mass, the patch size of $\patchsize$ that we use for training does not have enough high-flux or massive clusters for reliable training. 
This, however, is not a major obstacle for the current or future cosmological analysis. 
Note that the power spectrum and CMB lensing analyses require these extreme sources to be either masked or inpainted to avoid the challenges in properly modeling the artifacts due to filtering and biases in lensing at these locations.\\

\noindent
{\it Dependence on the training samples:} 
Finally, any generative model is only as good as the training data. 
In this work, we have not attempted to generate a new set of cosmological simulations from scratch but only to increase the number of realizations using existing simulations with the \diffusionmodel{} framework.
Our training data comes from the \agora{} simulation \citep{omori24} which adopted a specific set of cosmological and astrophysical parameters. 
Any mismatches between these assumptions and the true sky could lead to biases when the generative model is applied to data. 
Whether these differences are negligible depends on the specific application. 
Future work could explore training on an ensemble of simulations with varied physical assumptions, or conditioning on cosmological parameters to marginalize over modeling uncertainties.

\subsection{Comparison with other generative modeling approaches}
Generative models such as Generative Adversarial Networks (GANs), Variational Autoencoders (VAEs), Normalizing Flows, and Denoising Diffusion Probabilistic Models (\diffusionmodel{}s) offer distinct trade-offs for modeling extragalactic foregrounds.
GANs have been used to model astrophysical sources, but are notoriously unstable during training due to issues like mode collapse and vanishing gradients \citep{dhariwal21}. Additionally, GANs do not provide an explicit likelihood \citep{grover18}, limiting their applicability in Bayesian inference frameworks. 
VAEs provide a more stable training process and offer an explicit likelihood-based framework, making them attractive for Bayesian applications. However, they often rely on simplistic latent variable distributions (e.g., Gaussian), which can result in oversmoothed outputs and loss of angular resolution, which is particularly problematic for fine-scale structures in astrophysical maps \citep{thorne17}.
Normalizing Flows allow for exact likelihood estimation and efficient sampling via invertible mappings, but their expressiveness is limited by topological constraints imposed by the choice of priors, especially when modeling multi-modal distributions \citep{zhang21}.
\diffusionmodel{}s provide stable training procedures without mode collapse, producing diverse and robust samples without the need for manual intervention or tweaking. 
\diffusionmodel{}s can preserve the small-scale features present in astrophysical foregrounds, such as the resolved dusty galaxies (CIB) and galaxy clusters (tSZ). Recent comparative studies demonstrate that diffusion models outperform GANs and VAEs in tasks such as denoising weak lensing mass maps and reconstructing non-Gaussian lensing signals. For example, \citet{aoyama25} found that diffusion models achieve more accurate recovery of cosmological statistics across multiple benchmarks, although at the cost of increased training time and memory usage.

\section{Conclusions}
\label{sec_conclusion}
In this work, we have demonstrated the application of Denoising Diffusion Probabilistic Models to synthesize realistic extragalactic foreground maps, focusing on the Cosmic Infrared Background and thermal Sunyaev–Zel’dovich maps at 150 GHz from the \agora{} suite of simulations. 
We showed that the \diffusionmodel{} learns the full joint distribution of these signals, enabling the generation of correlated sky patches that faithfully reproduce not only the two-point (power spectrum) statistics but also higher-order non-Gaussian features such as the bispectrum, trispectrum, pixel intensity distributions, and Minkowski functionals. 

While \diffusionmodel{}s produce high-fidelity samples, we identified key challenges in their practical deployment. Notably, rare high-intensity pixels that can contribute significantly to the Poisson part of the power spectrum are underrepresented in the generated maps.
This limitation leads to a slight deficit in the amplitude of the generated power spectra. 
We found that a simple post-hoc variance rescaling of the generated maps effectively corrects this issue, improving agreement across summary statistics. 
A more significant challenge lies in scaling these models to larger sky areas. The number of available training samples decreases with increasing patch size, since only a handful of correlated multi-tracer simulations currently exist. Additionally, the fidelity of the \diffusionmodel{}s is tied to the quality and assumptions of the training data. Any discrepancies between these assumptions and the true sky could introduce modeling biases. These limitations must be addressed before broader deployment of \diffusionmodel{}s in cosmological analyses.

The \diffusionmodel{} framework naturally generalizes to additional foreground components, and can be extended across multiple frequencies to model full spectral energy distributions. This makes \diffusionmodel{}s a promising tool for forward modeling in CMB pipelines, where fast and accurate simulations are needed for foreground marginalization and unbiased parameter estimation. Physics-based simulations like \agora{} require thousands of computing hours to produce a single realization, while \diffusionmodel{}s can generate new CIB–tSZ map pairs in seconds. This opens the door to large-scale, rapid generation of realistic foreground realizations, something that has been out of reach with traditional simulation methods.
We share our code and the plotting scripts in this \href{https://github.com/Karthikprabhu22/diffusion_model}{GitHub repo$^{\text{\faGithub}}$}.



\section*{Acknowledgments}
We would like to thank Marius Millea for useful discussions and Yuuki Omori for consultation about \agora{} \citep{omori24} data products.
KP and LK acknowledge the support of Michael and Ester Vaida and the National Science Foundation via award OPP-1852617. 
SR acknowledges the support by the Illinois Survey Science Fellowship from the Center for AstroPhysical Surveys at the National Center for Supercomputing Applications.

The \agora{} simulations use the CosmoSim database, which is a service by the Leibniz-Institute for Astrophysics Potsdam (AIP).
The MultiDark database was developed in cooperation with the Spanish MultiDark Consolider Project CSD2009-00064.

This work made use of the following computing resources: Illinois Campus Cluster, a computing resource that is operated by the Illinois Campus Cluster Program (ICCP) in conjunction with the National Center for Supercomputing Applications (NCSA) and which is supported by funds from the University of Illinois at Urbana Champaign; the computing resources provided on Crossover, a high-performance computing cluster operated by the Laboratory  Computing Resource Center at Argonne National Laboratory; and the National Energy Research Scientific Computing Center (NERSC), a DOE Office of Science User Facility supported by the Office of Science of the U.S. Department of Energy under Contract No. DE-AC02-05CH11231.


\appendix

\section{Training Details}
\label{appendix_training}
The training dataset comprises roughly $\sim$4,000 patches of size $256 \times 256 \times 2$ (CIB + tSZ channels). We use a U-Net architecture with four encoder and decoder blocks, with channel depths of 64, 128, 256, and 512. The network includes sinusoidal timestep embeddings and self-attention layers at intermediate resolutions. A detailed breakdown of the architecture is provided in Table~\ref{tab_unet}. 

Training is carried out over $T = 1000$ diffusion steps using a sigmoid noise schedule, with $\beta_1 = 10^{-4}$ and $\beta_{1000} = 10^{-2}$. We optimize the velocity-prediction objective as defined in \citet{salimans22}, using the Adam optimizer \cite{kingma17} with a learning rate of $1 \times 10^{-4}$ and a batch size of 16. The model is trained for approximately 100,000 steps on a single NVIDIA A100 GPU, with checkpointing enabled. Full training takes around 30 hours. Once trained, sampling takes roughly 1–2 seconds per patch on the same hardware.

\begin{table}[ht]
\centering
\begin{tabular}{lll}
\toprule
\textbf{Stage} & \textbf{Output\,($H\times W\times C$)} & \textbf{Key operations} \\
\midrule
Input conv      & $256\times256\times64$  & $7\times7$ Conv (stride 1, pad 3) \\
Time embedding  & --                      & Sinusoidal pos‑emb $\to$ Linear $\to$ GELU $\to$ Linear \\
Down 0          & $128\times128\times64$  & $2\times$ ResNet + LinearAttn + Downsample \\
Down 1          & $64\times64\times128$   & $2\times$ ResNet + LinearAttn + Downsample \\
Down 2          & $32\times32\times256$   & $2\times$ ResNet + LinearAttn + Downsample \\
Down 3          & $32\times32\times512$   & $2\times$ ResNet + Full Attn + $3\times3$ Conv \\
Bottleneck      & $32\times32\times512$   & ResNet $\to$ Full Attn $\to$ ResNet \\
Up 3            & $64\times64\times256$   & $2\times$ ResNet + Full Attn + Upsample \\
Up 2            & $128\times128\times128$ & $2\times$ ResNet + LinearAttn + Upsample \\
Up 1            & $256\times256\times64$  & $2\times$ ResNet+ LinearAttn + Upsample \\
Up 0            & $256\times256\times64$  & $2\times$ ResNet + LinearAttn + $3\times3$ Conv \\
Final           & $256\times256\times2$   & ResNet (skip) + $1\times1$ Conv \\
\midrule
\multicolumn{2}{l}{\textbf{Total trainable parameters}} & 35.7 M \\
\bottomrule
\end{tabular}
\caption{U-Net architecture used in this work. Input maps are of shape $256\times256\times2$, representing CIB+tSZ channels. Total trainable parameters: 35.7 million.}
\label{tab_unet}
\end{table}

\section{Correlations across frequency}
\label{appendix_multifrequency}
\begin{figure*}
    \centering
    \includegraphics[width=\linewidth, keepaspectratio]{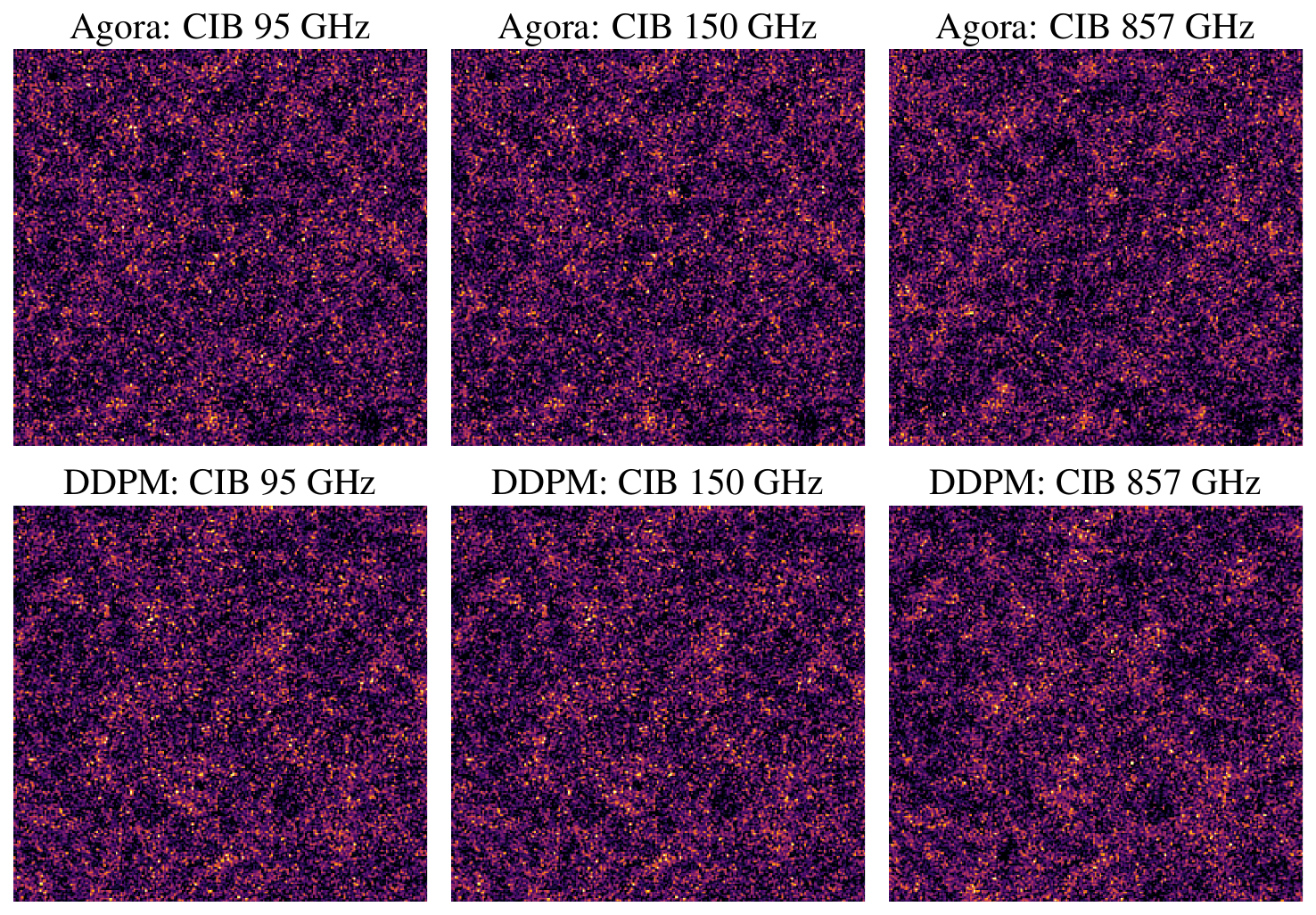}
    \caption{Examples of CIB maps at 95, 150, and 857 GHz from the test set  (top row) and the \diffusionmodel{} (bottom row). Each row shows the same sky patch across different frequencies. The \diffusionmodel{}-generated maps preserve the highly correlated filamentary structures and small-scale fluctuations present in the \agora{} simulation. For visualization, each map is standardized by subtracting its mean and dividing by its standard deviation to ensure perceptual consistency across channels. 
}
    \label{fig_example_cib_triplet}
\end{figure*}

\begin{figure*}
    \centering
\includegraphics[width=\linewidth, keepaspectratio]{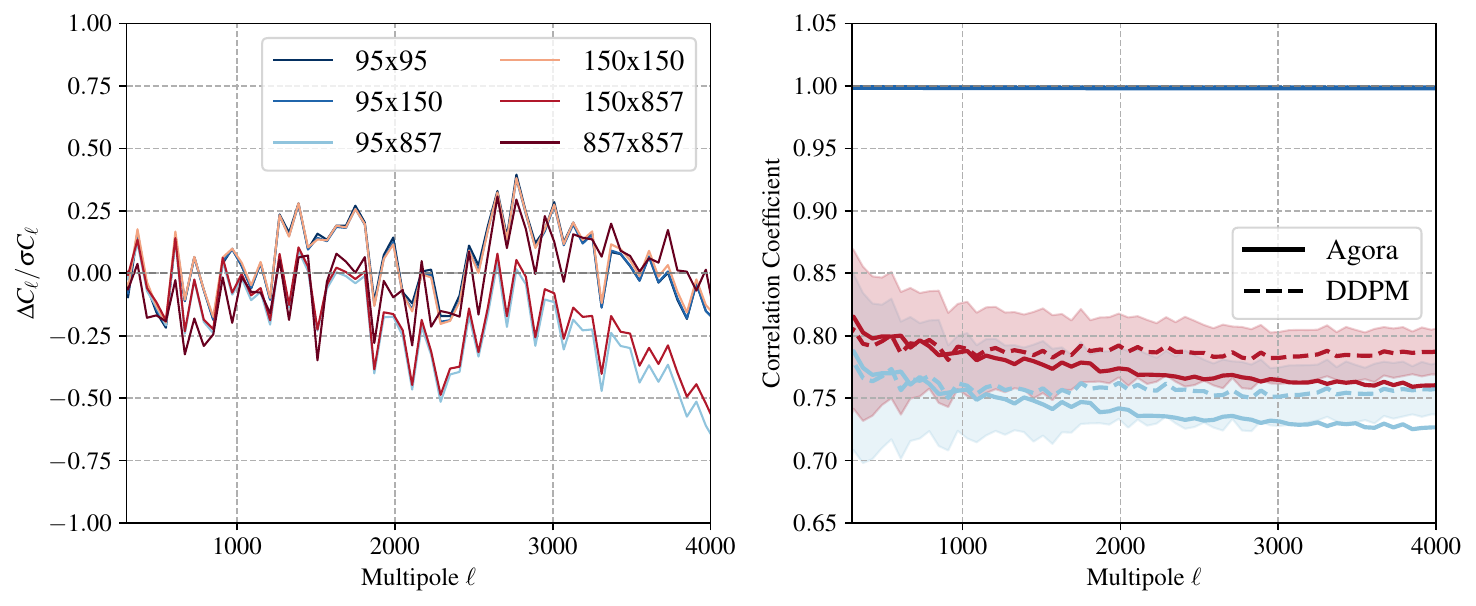}
    \caption{
    \refresponse{{\it Left panel:} Standard errors of the \diffusionmodel{} power spectra compared to the \agora{} test set for all auto- and cross-power spectra of the 95, 150, and 857 GHz CIB maps, averaged over 200 samples. The errors in each spectra is generally within $50\%$. {\it Right Panel:} Cross-correlation coefficients between different frequency bands computed from the \agora{} test maps (solid) and \diffusionmodel{} samples (dashed). The \diffusionmodel{} accurately reproduces the correlations observed in the \agora{} simulations. }
}
    \label{fig_example_cib_triplet_PS}
\end{figure*}

To assess the ability of \diffusionmodel{}s to capture the inter-frequency correlations, we trained a model to jointly generate CIB maps at 95, 150, and 857 GHz. 
The CIB fields at 95 and 150 GHz are expected to be highly correlated, while decorrelation becomes apparent at 857 GHz due to the increased contribution from lower-redshift sources.  
Figure~\ref{fig_example_cib_triplet} demonstrates that the \diffusionmodel{} successfully reproduces the correlated structures present in the CIB maps across three frequencies. 
In addition to visual inspection, we quantitatively assess the model’s ability to reproduce the frequency-dependent correlations using angular power spectra. 
\refresponse{In Figure~\ref{fig_example_cib_triplet_PS}, we show the standard errors between the \diffusionmodel{} and \agora{} power spectra for all auto- and cross-power spectra for the three frequencies. The errors are generally within 50\% of the standard deviation.} 
The right panel shows the cross-correlation coefficients between the 95, 150, and 857 GHz maps. 
The \diffusionmodel{} captures the near-unity correlation between 95 and 150 GHz and $\sim$ 80\% (77\%) between 150 (95) GHz and 857 GHz, consistent with the \agora{} simulations. 
\refresponse{We observe that the decorrelation between frequency bands is slightly milder for the DDPM samples at higher multipoles compared to the original \agora{} simulations. 
Similar to the investigation done in \S~\ref{sec_bispectrum_trispectrum}, we mask the extreme-value pixels in the CIB maps and recalculate the correlation coefficients for both \agora{} and \diffusionmodel. 
The impact of this masking is only marginal and well within the errors shown in Fig.~\ref{fig_example_cib_triplet}. 
We also defer a careful study of this behavior to a future work.
}

\section{Higher order auto- and cross- moments}
\label{appendix_moments}
\begin{figure*}
    \centering
    \includegraphics[width=\linewidth, keepaspectratio]{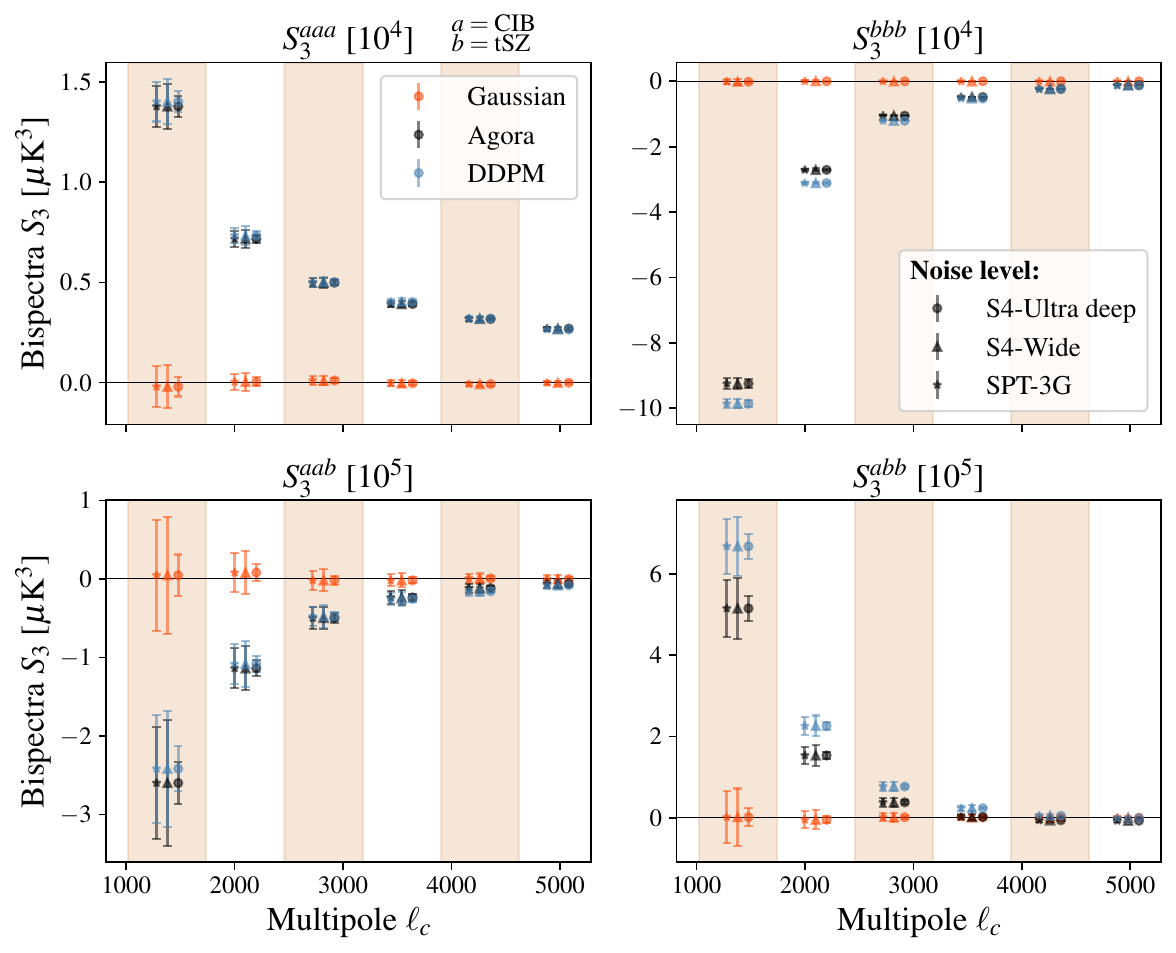}
    \caption{
    Joint bispectrum ($S_3$) statistics of CIB and tSZ maps computed from 800 patches of Gaussian, \agora{}, and \diffusionmodel{} samples. Each panel shows $S_3$ for different channel combinations: CIB-only ($S^{aaa}$), tSZ-only ($S^{bbb}$), and cross terms (e.g., $S^{aab}$ indicates two CIB and one tSZ input). Error bars denote the standard error across realizations under three instrumental noise levels: SPT-3G, S4-Wide, and S4-Ultra deep. Alternating shaded regions indicate multipole bands used for moment computation.
 }
    \label{fig_joint_bispectra}
\end{figure*}

\begin{figure*}
    \centering
    \includegraphics[width=\linewidth, keepaspectratio]{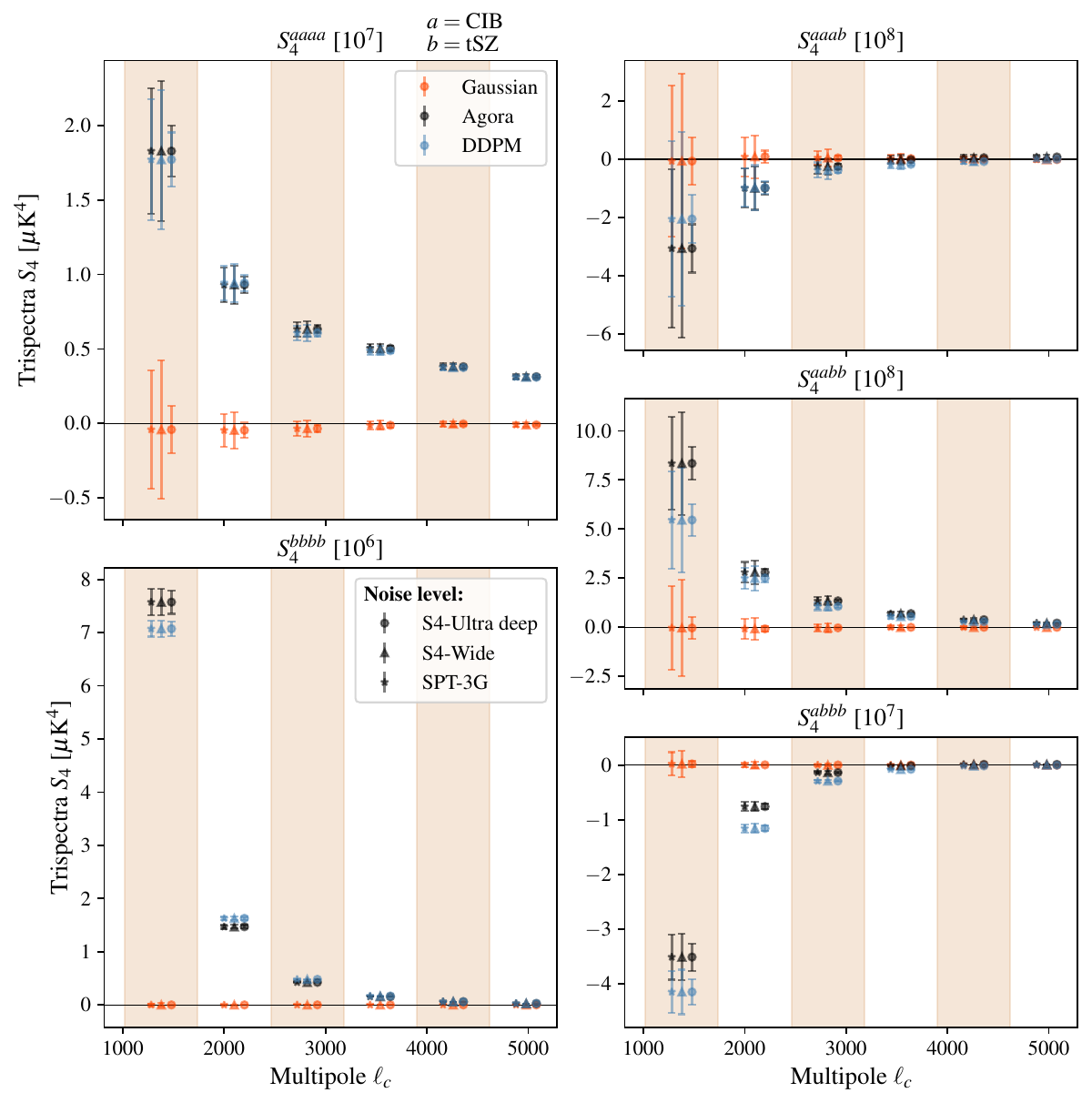}
    \caption{Joint trispectrum ($S_4$) statistics of CIB and tSZ maps, using the same format as Figure~\ref{fig_joint_bispectra}.
    }
    \label{fig_joint_trispectra}
\end{figure*}

In Figure~\ref{fig_joint_bispectra} and Figure~\ref{fig_joint_trispectra}, we show a more detailed version of Figure~\ref{fig_moments_joint}. We present the bispectra and trispectra for CIB-only, tSZ-only, and all possible CIB–tSZ combinations across three noise levels (SPT-3G, S4-Wide, and S4-Ultra deep). 

\ifdefined\jcapformat
    \bibliographystyle{apsrev4-1}
\else
    \bibliographystyle{aasjournal}
\fi
\bibliography{ddpm}
\end{document}